\pdfoutput=1
\documentclass[prl,aps,twocolumn,superscriptaddress,showpacs]{revtex4-1}

\usepackage{graphicx}
\usepackage{epstopdf}
\usepackage{amsmath}

\begin{document}

\title{Weak spin interactions in Mott insulating La$_{2}$O$_{2}$Fe$_{2}$OSe$_{2}$}

\author{E. E. McCabe}
\affiliation{Department of Chemistry, Durham University, Durham DH1 3LE, UK}
\affiliation{School of Physical Sciences, University of Kent, Canterbury, CT2 7NH, UK}
\author{C. Stock}
\affiliation{School of Physics and Astronomy, University of Edinburgh, Edinburgh EH9 3JZ, UK}
\author{E. E. Rodriguez}
\affiliation{Department of Chemistry of Biochemistry, University of Maryland, College Park, MD, 20742, U.S.A.}
\author{A. S. Wills}
\affiliation{Department of Chemistry, University College London, 20 Gordon Street, London WC1H 0AJ, UK}
\author{J. W. Taylor}	
\affiliation{ISIS Facility, Rutherford Appleton Labs, Chilton Didcot, OX11 0QX, UK}
\author{J. S. O. Evans}
\affiliation{Department of Chemistry, Durham University, Durham DH1 3LE, UK}
\date{\today}

\begin{abstract}

Identifying and characterizing the parent phases of iron-based superconductors is an important step towards understanding the mechanism for their high temperature superconductivity.  We present an investigation into the magnetic interactions in the Mott insulator La$_{2}$O$_{2}$Fe$_{2}$OSe$_{2}$. This iron oxyselenide adopts a 2-$k$ magnetic structure with low levels of magnetic frustration. This magnetic ground state is found to be dominated by next-nearest neighbor interactions $J_{2}$ and $J_{2 '}$ and the magnetocrystalline anisotropy of the Fe$^{2+}$ site, leading to 2D-Ising-like spin $S$=2 fluctuations.  In contrast to calculations, the values are small and confine the spin excitations below $\sim$ 25 meV.  This is further corroborated by sum rules of neutron scattering. This indicates that superconductivity in related materials may derive from a weakly coupled and unfrustrated magnetic structure. 

\end{abstract}

\maketitle

The discovery of iron-based superconductivity at high temperatures in pnictide~\cite{Kamihara08:130} and chalcogenide~\cite{Margadonna09:80} systems highlights the importance of magnetism in high-$T_{c}$ superconductivity.~\cite{Cruz08:452} Despite the similar phase diagrams and the proximity of magnetism to superconductivity reported for both the cuprate and iron-based superconductors, these materials otherwise seem remarkably different: the cuprate systems are based on doping a strongly-correlated Mott insulating state,~\cite{Lee06:78} while the parent phases for the iron-based materials are either metallic, semiconducting, or semimetallic.~\cite{Mazin10:47,Johnston10:59,Paglione10:6}  However, recent work has revealed electron correlation effects in iron pnictides suggesting that the iron-based systems may be close to the Mott boundary, yet a strongly correlated parent compound has not been clearly identified for chalcogenide and pnictide superconductors.~\cite{Si09:629,Qazilbash09:5}  Also, the spin state of the Fe$^{2+}$ in these systems is not understood with different theories suggesting $S$=1 or 2 ground states~\cite{Yin11:10,Si08:101,Kruger09:79}.  In this paper, we investigate the magnetic interactions in the Mott insulating iron oxyselenide La$_{2}$O$_{2}$Fe$_{2}$OSe$_{2}$. 

\begin{figure}[t] 
\includegraphics[width=1.0\linewidth,angle=0.0]{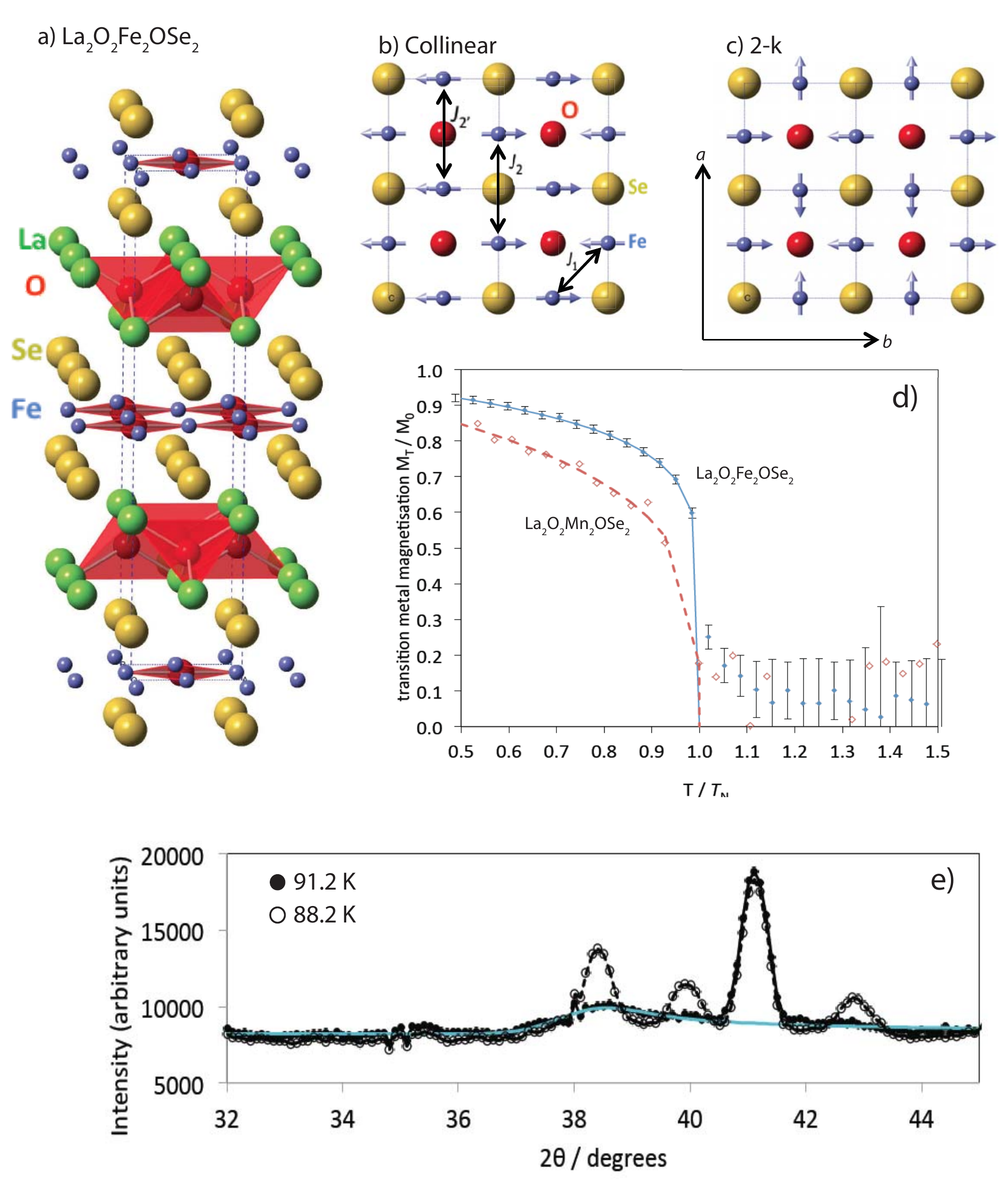}
\caption{[color online]  a) nuclear cell of La$_{2}$O$_{2}$Fe$_{2}$OSe$_{2}$ b) collinear model and c) 2-$k$ model with the three intraplanar exchange interactions $J_{1}$, $J_{2}$ and $J_{2 '}$ shown; d) evolution of magnetic moment for La$_{2}$O$_{2}$Fe$_{2}$OSe$_{2}$ and La$_{2}$O$_{2}$Mn$_{2}$OSe$_{2}$ (Ref. \onlinecite{Free11:23}) with with $M_{0Fe}$ = 3.701(8) $\mu_{B}$, $T_{\mathrm{N}}$=89.50(3) K and  $\beta_{Fe}$=0.122(1);  $M_{0Mn}$ = 4.5(2) $\mu_{B}$, $T_{\mathrm{N}}$ = 168.1(1) K and $\beta_{Mn}$ = 0.24(3) and e) shows narrow 2$\theta$ range of raw NPD data for La$_{2}$O$_{2}$Fe$_{2}$OSe$_{2}$ collected at 91.2 K and at 88.2 K, the Warren-type peak shown by solid blue line.}
\label{structure}
\end{figure}

This layered material (Fig. \ref{structure}$a$) adopts a tetragonal crystal structure composed of fluorite-like [La$_{2}$O$_{2}$]$^{2+}$ layers and [Fe$_{2}$O]$^{2+}$ sheets that are separated by Se$^{2-}$ anions. The [Fe$_{2}$O]$^{2+}$ sheets adopt an unusual anti-CuO$_{2}$ arrangement with Fe$^{2+}$ cations coordinated by two in-plane oxygens and four Se$^{2-}$ anions above and below the plane, leading to layers of face-shared FeO$_{2}$Se$_{4}$ trans octahedra.~\cite{Mayer92:31}  The Fe grid is similar to that in LaFeAsO and FeSe, has similar $\sim$ 90$^{\circ}$ Fe-Se-Fe interactions but contains additional in-plane O$^{2-}$ ions.

La$_{2}$O$_{2}$Fe$_{2}$OSe$_{2}$ has been described as a Mott insulator and theoretical work suggests that it is more strongly correlated than LaFeAsO.~\cite{Zhu10:104}  La$_{2}$O$_{2}$Fe$_{2}$OSe$_{2}$ orders antiferromagnetically (AFM) below $\sim$ 90 K~\cite{Free10:81} and two magnetic structures have been discussed for the [Fe$_{2}$O]$^{2+}$ layers: a collinear model (Fig. \ref{structure}$b$) similar to that reported for Fe$_{1+x}$Te;~\cite{Bao09:102,Rodriguez11:84} and the 2-$k$ model (Fig. \ref{structure}$c$) first proposed for Nd$_{2}$O$_{2}$Fe$_{2}$OSe$_{2}$.~\cite{Fuwa10:22}  These two models are indistinguishable from powder diffraction work and in the absence of single crystals of sufficient size and quality, this ambiguity has not been resolved. We present experimental results here that favour the 2-$k$ model proposed by Fuwa et al~\cite{Fuwa10:22} and hope to resolve this ambiguity.

The related pnictide and chalcogenide parent compounds have been the subject of theoretical and experimental studies. Analogous to the cuprates, the spin exchange constants and spin-wave dispersions in these parent compounds are large, extending up to energy transfers of $\sim$ 100 meV, reflecting strong Fe-Fe coupling.~\cite{Coldea01:86,Stock07:75,Ewings08:78,Zhao09:5,Diallo09:102,Lipscombe11:106} Electronic structure calculations for La$_{2}$O$_{2}$Fe$_{2}$OSe$_{2}$ suggest similar exchange constants to the pnictides but with considerable electronic band narrowing ~\cite{Zhu10:104}. Until now, neutron inelastic measurements to corroborate such predictions have not been reported for La$_{2}$O$_{2}$Fe$_{2}$OSe$_{2}$.
We present a combined study of the magnetic structure and fluctuations to understand the interactions in La$_{2}$O$_{2}$Fe$_{2}$OSe$_{2}$ using neutron powder diffraction (NPD) and inelastic measurements.  Full experimental details are provided in the supplementary information.  

We first discuss the elastic magnetic scattering near $T_{\mathrm{N}}$ ($\sim$ 89 K) (Fig. \ref{structure}$e$).  A broad, low intensity, asymmetric Warren-like peak develops between 103 and 91 K centered at $\sim$37$^{\circ}$ 2$\theta$, characteristic of 2D short-ranged ordering.~\cite{Warren41:59} Fitting with a Warren function gives a 2D correlation length of $\sim$ 23 \AA\ at 103 K that increases to $\sim$ 90 \AA\ (about 20 times the in-plane cell parameter) just above $T_{\mathrm{N}}$.  Below $T_{\mathrm{N}}$, magnetic Bragg reflections appear with the most intense peak at 2$\theta \sim$ 38$^{\circ}$, such that any remaining diffuse scatter becomes hard to fit.

\begin{figure}[t]
\includegraphics[width=1.0\linewidth,angle=0.0]{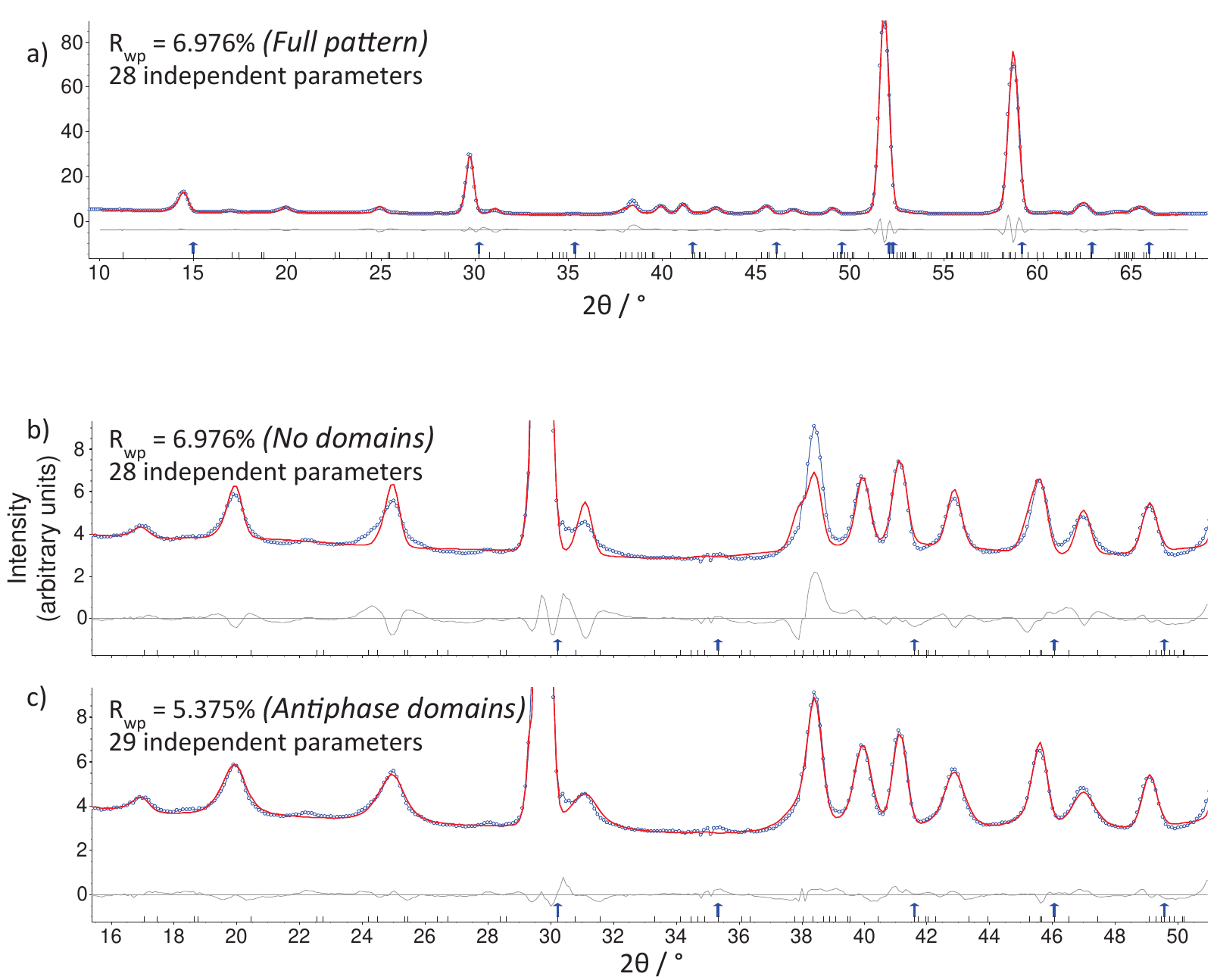}
\caption{[color online] Rietveld refinements (D20, $\lambda$=2.41 \AA) with the 2-$k$ model showing a) wide 2$\theta$ range with both nuclear (blue arrows) and magnetic (black tick marks) phases. $b)$ refinement with the \textit{same} peak shape for both nuclear and magnetic phases;  $c)$ refinement including antiphase boundaries in the magnetic phase. Observed and calculated (upper) and difference (lower, at zero intensity) profiles are shown by blue points, red, and grey lines, respectively.  The tick marks do not include a refined zero offset of $\sim$ 0.4$^{\circ}$.}
\label{diffraction}
\end{figure}

Magnetic Bragg reflections appear below $T_{\mathrm{N}}$, to which the 2-$k$ (Fig. \ref{diffraction}) and collinear spin models give indistinguishable fits.  In contrast to the report on Sr$_{2}$F$_{2}$Fe$_{2}$OS$_{2}$, there is no difference in the magnitude of Fe moments for these two models.~\cite{Zhao13:87}  The magnetic Bragg reflections observed for La$_{2}$O$_{2}$Fe$_{2}$OSe$_{2}$ are anisotropically broadened similar to Sr$_{2}$F$_{2}$Fe$_{2}$OS$_{2}$, suggesting that both have similar magnetic microstructures. This peak broadening can be described by an expression for antiphase boundaries perpendicular to the $c$-axis~\cite{Her07:63} (Fig. \ref{diffraction}$c$) with a magnetic correlation length $\xi_{c}(T=2 K)$=45(3)\AA\ that is essentially independent of temperature ($\xi_{c} (T=88 K)$=42(6)\AA\ ).  No such peak broadening has been reported for the Mn$^{2+}$ and Co$^{2+}$ analogues.~\cite{Ni10:82,Fuwa10:132,Free11:23}  

Sequential NPD Rietveld refinements indicate a smooth increase in the ordered Fe$^{2+}$ moment on cooling. This magnetic order parameter is shown in Fig. \ref{structure}$d$) with critical exponent $\beta_{Fe}$=0.122(1), similar to the 2D-Ising like behavior of La$_{2}$O$_{2}$Co$_{2}$OSe$_{2}$ and BaFe$_{2}$As$_{2}$.~\cite{Fuwa10:132,Wilson10:81}. This is in contrast to the Mn analogue with an exponent $\beta$=0.24(3) (Fig. \ref{structure}$d$)) reflecting greater 3D-like character.~\cite{Free11:23} The ordered Fe$^{2+}$ moment in La$_{2}$O$_{2}$Fe$_{2}$OSe$_{2}$ determined from our Rietveld refinements (3.50(5) $\mu_{B}$ at 2 K) is larger than that reported previously ($\sim$2.8 $\mu_{B}$)~\cite{Free10:81,Ni11:83}, due to improved fitting of magnetic Bragg peaks (Fig. \ref{diffraction}$c$); our value is similar to that reported for Sr$_{2}$F$_{2}$Fe$_{2}$OS$_{2}$ (3.3(1) $\mu_{B}$)~\cite{Zhao13:87} and in the parent phase of superconducting K$_{x}$Fe$_{2-y}$Se$_{2}$ (3.3 $\mu_{B}$).~\cite{Bao11:28,Zhao12:109}

\begin{figure}[t]
\includegraphics[width=0.9\linewidth,angle=0.0]{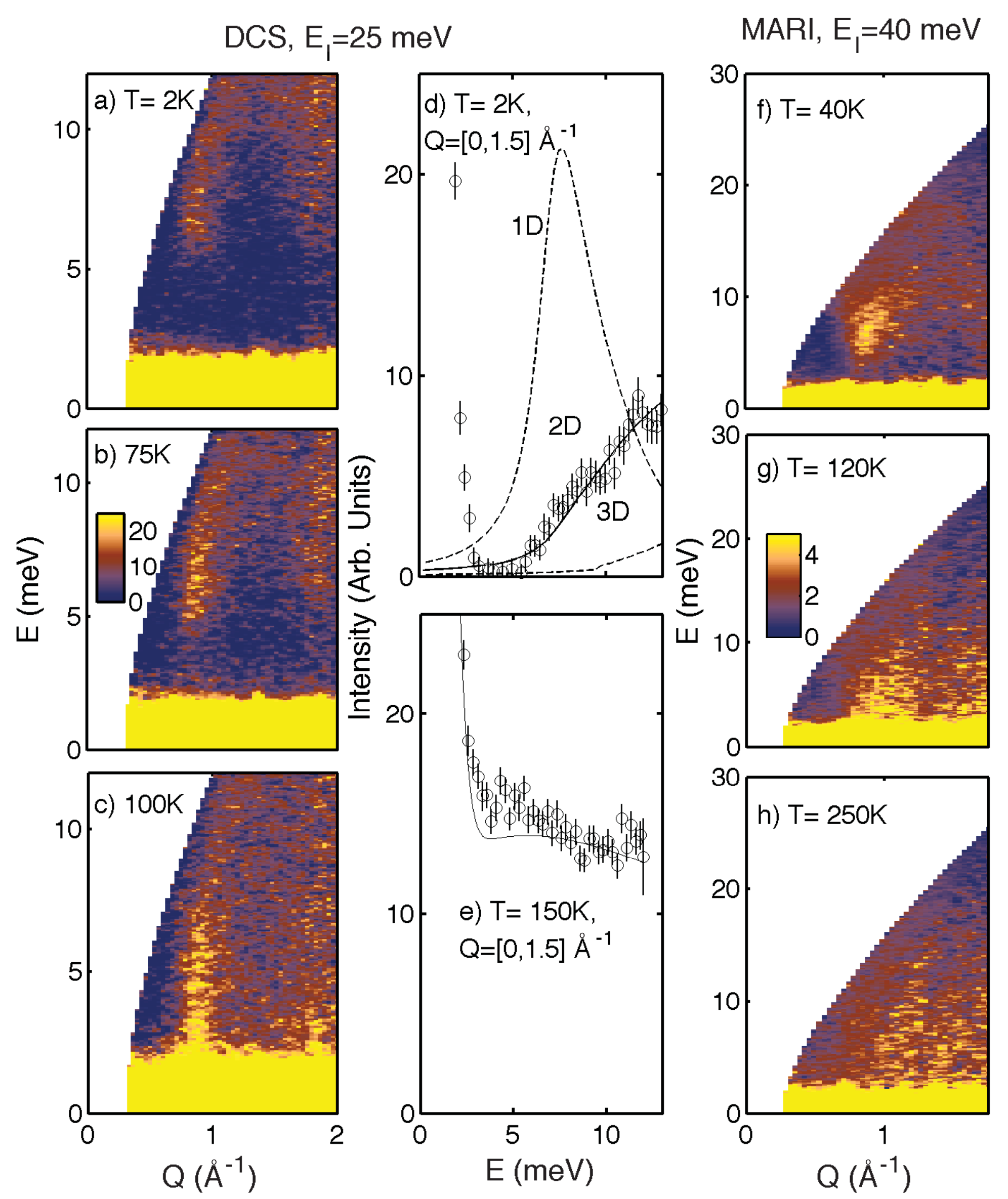}
\caption{[color online]  a-c) powder-averaged spectra measured on DCS; d) shows the momentum-integrated energy scan at 2 K (upper) and 150 K (lower), the curves are calculations using a single-mode analysis with a 1D model, a 2D model and a 3D model; f-h) plot the powder-averaged temperature spectra taken on the MARI spectrometer.}
\label{inelastic_temp}
\end{figure}

We now discuss spin excitations characterizing the magnetic interactions shown in Fig. \ref{structure}.  Fig. \ref{inelastic_temp} shows the temperature-dependent, powder-averaged inelastic response. The spectra at 2 K show the magnetic response is gapped and localized in momentum (Fig. \ref{inelastic_temp}$a$) and softens on warming (Fig. \ref{inelastic_temp}$b$) until gapless scattering is observed for $T>T_{\mathrm{N}}$ (Fig. \ref{inelastic_temp}$c$). This is further illustrated in Fig. \ref{inelastic_temp}$d$ and $e)$ (showing $Q$-integrated energy scans) and in lower resolution scans $f-g$. The intensity distribution at the gap edge is sensitive to the dimensionality of the interactions and can be quantified through use of the first moment sum rule. Fig. \ref{inelastic_temp}$d$ shows a comparison of the momentum integrated intensity with calculations based on the single mode approximation for an isotropic dispersion in a one-dimensional (1D) chain, 2D plane or 3D structure.~\cite{Hohenberg74:10, Hammar98:57, Stock09:103} The 2D model gives the best description consistent with the 2D-Ising critical properties discussed above.

Scans that probe larger energy transfers are shown in Figs. \ref{inelastic_temp}$f-h$. Surprisingly, the magnetic excitations extend up to only $\sim$ 25 meV.  This small band accounts for all of the expected spectral weight, confirmed by integrating the intensity and comparing with the zeroeth sum rule ($\tilde{I}={{\int d^{2}Q \int dE S(\vec{Q},E)} / {\int d^{3}Q}}=S(S+1)$).  Our inelastic data (over energy ranges shown in Fig. \ref{inelastic_temp}$e$) give $\tilde{I}_{inelastic}=$ 3.2(4) for the dynamic response. The elastic magnetic moment of 3.5 $\mu_{B}$ (determined from NPD discussed above) implies an elastic contribution to the above integral of $\tilde{I}_{static}$=2.7(1), giving $\tilde{I}$= 5.9(4), close to the $S$=2 value of 6.   Over this narrow energy range, all magnetic spectral weight is accounted for. 

\begin{figure}[t]
\includegraphics[width=0.7\linewidth,angle=0.0]{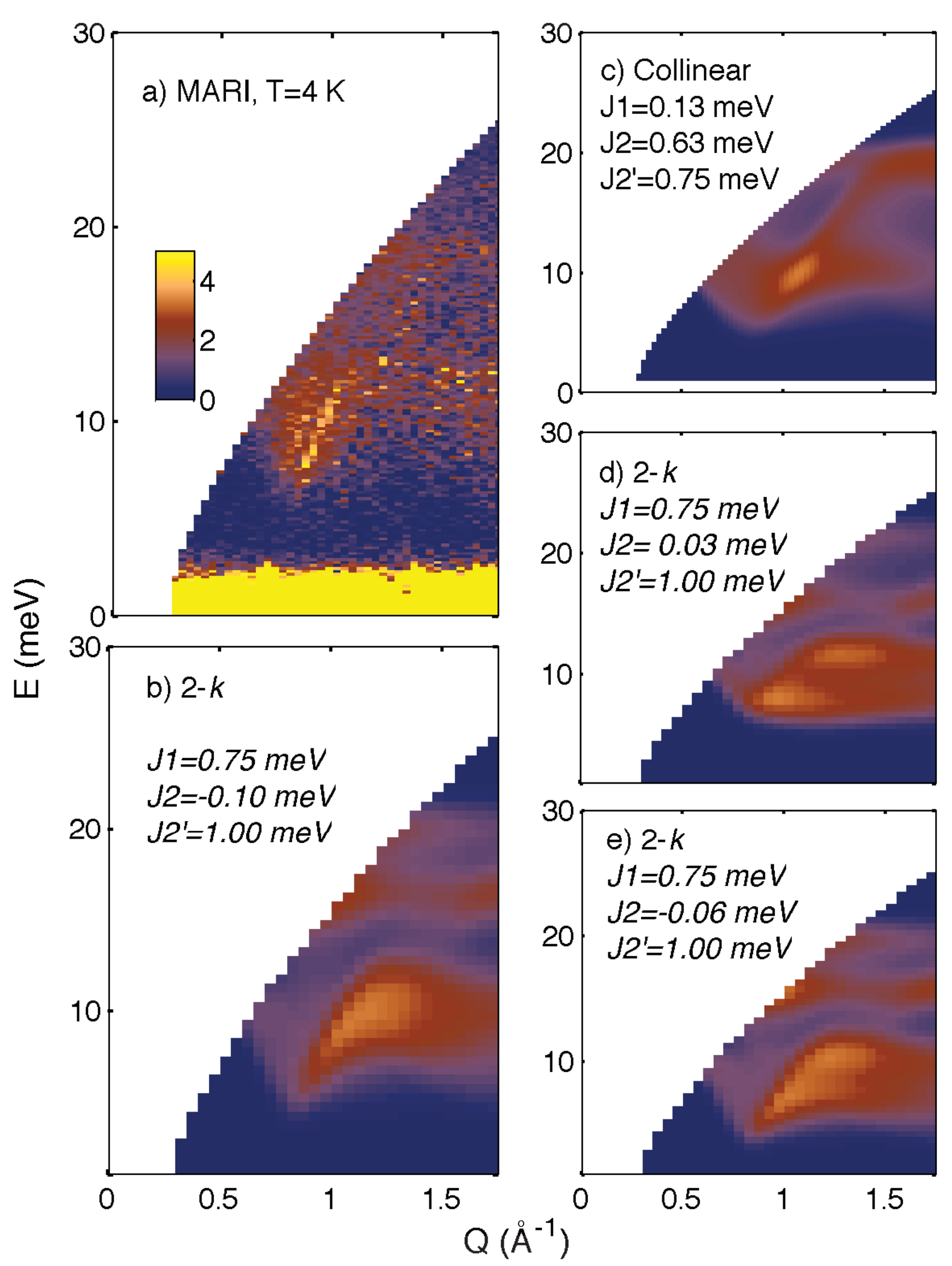}
\caption{[color online]  a) MARI scans with E$_{i}$ = 40 meV and spin-wave models for b) the 2-$k$ structure and c) the collinear magnetic structure; d) and e) show the effect of weak AFM and FM values of the J$_{2}$ exchange interaction on simulated spectra.}
\label{inelastic_model}
\end{figure}

This analysis demonstrates that the total bandwidth of the spin excitations is only $\sim$20 meV.  This is remarkably small when compared with Mott insulating La$_{2}$CuO$_{4}$ and YBa$_{2}$Cu$_{3}$O$_{6+x}$ (with a bandwidth of over 300 meV) and with the parent phases of the pnictides (the top of the band in BaFe$_{2}$As$_{2}$ is $\sim$100 meV and $\sim$150 meV in CaFe$_{2}$As$_{2}$) or the chalcogenide Fe$_{1+x}$Te (where excitations extend up to $\sim$150-200 meV).~\cite{Coldea01:86,Stock07:75,Diallo09:102,Lipscombe11:106,Ewings08:78}  The small bandwidth observed for La$_{2}$O$_{2}$Fe$_{2}$OSe$_{2}$ implies that magnetic exchange interactions are about an order of magnitude smaller than in the cuprates and pnictides.

To estimate these exchanges, calculations were performed fixing the moment direction with a single-ion anisotropy and considering Heisenberg spin exchange.  The calculation is sensitive to the signs of the interactions and the ground state. These calculations were carried out based on both the collinear and 2-$k$ magnetic ground states (Fig. \ref{structure}) and results are shown in Fig. \ref{inelastic_model}. The experimental spectrum can be reproduced reasonably well for the 2-$k$ ground state with $J_{1}$ = 0.75 meV, $J_{2}$ = -0.10 meV and $J_{2 '}$ = 1.00 meV (Fig. \ref{inelastic_model}$b$) and for the collinear ground state with $J_{1}$ = 0.13 meV, $J_{2}$ = 0.63 meV and $J_{2 '}$ = 1.00 meV (Fig. \ref{inelastic_model}$c$). The predicted $\Theta_{CW}$, to be compared with a $T_{\mathrm{N}}$ $\sim$ 90 K, are $\sim$ 110 K for the 2-$k$ and $\sim$ 75 K for the collinear models.  These two models give comparable descriptions of the data and differ mainly in the sign of the $J_{2 '}$ interaction with the 2-$k$ (collinear) ground state giving a FM (AFM) value.

We now compare the collinear and 2-$k$ models.   The collinear model (Fig. \ref{structure}$b$) is a single-$k$ model with $\vec{k}$ = (0 ${1\over 2}$ ${1\over 2}$). This $k$-vector splits the moments of the Fe site ($4c$ site in $I$4/$mmm$) into two orbits that order under separate irreducible representations (irreps) with the moments along the $b$ axis. The irreps and basis vectors involved are labelled $N_{2}^{+}(B_{3g})$ and $N_{1}^{-}(B_{2g})$ according to ISODISTORT~\cite{Campbell06:39}, and $\Gamma_{2} \psi_{1}$ and $\Gamma_{3} \psi_{2}$ following SARAh.~\cite{Wills00:276} In terms of energy, none of the three intraplanar exchange interactions are satisfied in the collinear structure, making it disfavoured on energetic grounds. As the mean fields experienced by the different orbits are orthogonal, they would order separately and so this model would also be disfavoured on entropic arguments.

The 2-$k$ model (Fig. \ref{structure}$c$) can, to a first approximation, be described by the spin Hamiltonian involving single-ion anisotropies and Heisenberg terms with AFM $J_{1}$ and $J_{2 '}$ and FM $J_{2}$, consistent with calculations~\cite{Zhao13:87} and with the values postulated here. The nearest neighbor exchange $J_{1}$ is thought to be AFM in all known $Ln_{2}$O$_{2}$$M_{2}$OSe$_{2}$ materials and dominates for La$_{2}$O$_{2}$Mn$_{2}$OSe$_{2}$.~\cite{Ni10:82,Free11:23,Koo12:324}  However, in the 2-$k$ model, the $J_{1}$ interactions are unimportant as nearest neighbor moments are perpendicular. Instead, it is the next nearest neighbor $J_{2}$ and $J_{2 '}$ that dominate. DFT calculations predict that $J_{2}$ via Se$^{2-}$ is FM for $M$=Fe, but AFM for $M$=Mn and Co, while $J_{2 '}$ (180$^{\circ}$ exchange via O$^{2-}$) is predicted to be AFM for all $M$.~\cite{Goodenough:book} The FM $J_{2}$ Fe-Se-Fe interactions, predicted by DFT, are consistent with the FM chain structure reported for Ce$_{2}$O$_{2}$FeSe$_{2}$.~\cite{Mccabe11:47} 2D exchange concomitant with magnetocrystalline anisotropy (due to partially unquenched orbital angular momentum) is likely to stabilize the 2-$k$ structure (and the $k$ = (${1\over 2}$ ${1\over 2}$ 0) structure reported for La$_{2}$O$_{2}$Co$_{2}$OSe$_{2}$).~\cite{Free11:23,Fuwa10:132}  This agrees with the Ising-like character suggested to constrain $M^{2+}$ moments to lie along perpendicular local axes within the $ab$ plane for $M$ = Fe, Co (i.e. along Fe-O bonds in La$_{2}$O$_{2}$Fe$_{2}$OSe$_{2}$). This anisotropy is not found in the high spin $M$ = Mn$^{2+}$ for which orbital angular momentum is zero and moments are oriented out of the $ab$ plane.~\cite{Free11:23,Ni10:82}   This anisotropy overrides $J_{1}$ and with FM $J_{2}$ and AFM $J_{2 '}$, favors the 2-$k$ over the collinear model. 

To stabilize 2-$k$ structures, energy terms beyond second order isotropic or antisymmetric exchange (Dzyaloshinski-Moriya) are required. Anisotropic exchange arising from spin anisotropy is able to introduce higher order terms that can stabilize combining the 2-$k$ components. In doing so, the $C_{4}$ rotational symmetry that relates the two $k$ vectors is reintroduced into the magnetic symmetry, constraining the moments of what were two independent orbits in the single-$k$ structure, to be equal in magnitude and related in-phase. This constraint causes the magnetic ordering to satisfy entropic requirements and the transition is second order as observed here by experiment.

While the 2-$k$ structure cannot be stabilized by second order spin terms alone, it is useful to explore the structure in terms of the interactions in Fig. \ref{structure}, which still embodies the two orbit structure of the single-$k$ model. In it, with no net $J_{1}$ nearest neighbour interactions, the 2-$k$ model can be thought of as two interpenetrating square sublattices, each described by one of the two $k$-vectors. Within each sublattice, $J_{2 '}$ coupling leads to AFM Fe-O-Fe stripes which are coupled by FM $J_{2}$ Fe-Se-Fe interactions. The 2-$k$ model (and the $k$ = (${1\over 2}$ ${1\over 2}$ 0) structure described for La$_{2}$O$_{2}$Co$_{2}$OSe$_{2}$) could result from dominant $J_{2 '}$ interactions where $J_{2 '}$ $>>$ $J_{1}$, $J_{2}$.  This exchange scenario would lead to a network of perpendicular quasi-1D AFM Fe-O-Fe chains. However, our experimental results indicate 2D-like magnetic exchange interactions making this quasi-1D scenario unlikely.

The 2-$k$ model can be compared with the magnetic ordering reported for Fe$_{1+x}$Te~\cite{Li09:79} which is also composed of two interpenetrating square sublattices.~\cite{Bao09:102,Rodriguez11:84} First, the origin of the anisotropy within each sublattice in Fe$_{1+x}$Te (i.e AFM interactions along $a_{T}$ and FM interactions along $b_{T}$ where $T$ subscript denotes tetragonal unit cell) is ascribed to orbital ordering, while in La$_{2}$O$_{2}$Fe$_{2}$OSe$_{2}$, the anisotropy within each single-$k$ sublattice is due to different exchange interactions along each direction. Second, the mechanism for coupling the two sublattices differs, with double exchange interactions proposed for metallic Fe$_{1+x}$Te~\cite{Turner09:80} being less likely for insulating La$_{2}$O$_{2}$Fe$_{2}$OSe$_{2}$. Rather, the strong spin-anisotropy observed supports a coupling by high order anisotropic exchange terms.

The observation of a Warren peak characteristic of short-range magnetic ordering only $\sim$ 14 K above $T_{\mathrm{N}}$ (in contrast to $\sim$ 140 K above $T_{\mathrm{N}}$ for La$_{2}$O$_{2}$Mn$_{2}$OSe$_{2}$)~\cite{Ni10:82} further supports the assignment of the (less frustrated) 2-$k$ rather than the collinear model. This is because the 2-$k$ structure diminishes the effects of $J_{1}$ and avoids frustration of $J_{2}$ and $J_{2 '}$. With both $J_{2}$ and $J_{2 '}$ satisfied, the 2-$k$ structure involves less frustration than in the Mn analogues. The anisotropic broadening of magnetic Bragg reflections suggests that there is only a small energy cost for disrupting the magnetic ordering along $c$ (e.g. introducing stacking faults or antiphase boundaries) giving a reduced magnetic correlation length in this direction. 


DFT calculations have supported the notion of large exchange constants in this material and related iron-based systems, in contrast with our experimental results.  Given that $J$ is proportional to $4t^{2}/U$,~\cite{MacDonald98:37} these small $J$ values determined experimentally suggest a small hopping integral $t$ for these oxychalcogenides, consistent with theoretical work which describes band narrowing in these materials.~\cite{Zhu10:104}   These small $J$ values imply that local bonding is more important than in related materials such as Fe$_{1+x}$Te and $Ln$FeAsO, and that La$_{2}$O$_{2}$Fe$_{2}$OSe$_{2}$ is a more correlated system than current DFT work suggests.

The integrated intensity over the small band width of excitations recovers the total moment for $S$=2.  While this is consistent with a large ordered moment, it implies that Fe$^{2+}$ is in a weak crystal field favoring, a Hund's rules population of the $d$-orbitals which contrasts with suggestions of a $S$=1 ground state from analysis of pnictide and chalcogenide superconductors.~\cite{Turner09:80,Haule09:11}  Our analysis, combined with the large ordered magnetic moments reported in K$_{x}$Fe$_{2-y}$Se$_{2}$, may indicate that the $S$=1 parent state may need to be reconsidered.

In conclusion, Mott-insulating La$_{2}$O$_{2}$Fe$_{2}$OSe$_{2}$ adopts a multi component 2-$k$ magnetic structure.  This structure is stabilized by AFM $J_{2 '}$ and FM $J_{2}$ interactions and the magnetocrystalline anisotropy of the Fe site and leads to 2D-Ising like spin fluctuations around the critical point.  Surprisingly, the magnetic exchange interactions are very small in comparison with related systems and also the Mott-insulating cuprates and an integrated intensity analysis implies a $S$=2 ground state. This may indicate additional localization in these $Ln_{2}$O$_{2}$$M_{2}$OSe$_{2}$ materials which has not yet been explored theoretically.

We acknowledge STFC, EPSRC (EP/J011533/1), RSE, and the NSF (DMR-0944772) for funding. We thank Emma Suard (ILL), Ross Stewart (ISIS), and Mark Green for assistance.


%

\end{document}


\title{Supplementary information describing the spin excitations and magnetic structure of La$_{2}$O$_{2}$Fe$_{2}$OSe$_{2}$}

\author{E. E. McCabe}
\affiliation{Department of Chemistry, Durham University, Durham DH1 3LE, UK}
\affiliation{School of Physical Sciences, University of Kent, Canterbury, CT2 7NH, UK}
\author{C. Stock}
\affiliation{School of Physics and Astronomy, University of Edinburgh, Edinburgh EH9 3JZ, UK}
\author{E. E. Rodriguez}
\affiliation{Department of Chemistry of Biochemistry, University of Maryland, College Park, MD, 20742, U.S.A.}
\author{A. S. Wills}
\affiliation{Department of Chemistry, University College London, 20 Gordon Street, London WC1H 0AJ, UK}
\author{J. W. Taylor}
\affiliation{ISIS Facility, Rutherford Appleton Labs, Chilton Didcot, OX11 0QX, UK}
\author{J. S. O. Evans}
\affiliation{Department of Chemistry, Durham University, Durham DH1 3LE, UK}
\date{\today}

\begin{abstract}

Supplementary information regarding neutron elastic and inelastic scattering on La$_{2}$O$_{2}$Fe$_{2}$OSe$_{2}$ is presented.  Further details of the magnetic and nuclear structural refinements are given along with additional information pertaining to the analysis of the magnetic excitations discussed in the main text.  A description of the heuristic spin-wave Hamiltonian and model used to parameterize the spin excitations are given along with the calculated Curie-Weiss constants stated in the main text.

\end{abstract}

\maketitle

\section{Experimental details and sample details}

A 5.25 g sample of La$_{2}$O$_{2}$Fe$_{2}$OSe$_{2}$ was prepared following the method described by Free and Evans.~\cite{Free10:81}  Neutron powder diffraction data were collected on the high flux diffractometer D20 at the ILL with neutron wavelength ~2.41 \AA. The powder was placed in an 8 mm diameter cylindrical vanadium can (to a height of ~4 cm) and data were collected from 5-130$^{\circ}$ in $2\theta$. A 30 minute scan was carried out at 216 K, then 5 minute scans were collected on cooling at $\sim$ 3 K intervals, followed by a 40 minute scan at 2.4 K. Rietveld refinements were performed using TopasAcademic software. The diffractometer zero point and neutron wavelength were initially refined while sample cell parameters were fixed to values obtained at the same temperature from refinements using the previously published HRPD data.~\cite{Free10:81} The values for zero point and wavelength were then fixed at these values for subsequent refinements. Typically, a background was refined for each refinement, as well as unit cell parameters, atomic positions and a Caglioti peak shape. TopasAcademic permits Ònuclear-onlyÓ and Òmagnetic-onlyÓ phases to be included in refinements and the unit cell parameters of the magnetic phase were constrained to be integer multiples of those of the nuclear phase. The scale factor scales with the square of the unit cell volume; the scale factor for the nuclear phase was refined and that for the magnetic phase (with cell volume four times that of the nuclear phase) was constrained to be 0.0625 $\times$ that of the nuclear phase.  The web-based ISODISTORT software was used to obtain a magnetic symmetry mode description of the magnetic structure; magnetic symmetry modes were then refined corresponding to the two magnetic structures.

For inelastic neutron scattering measurements, the 5.25 g of La$_{2}$O$_{2}$Fe$_{2}$OSe$_{2}$ prepared here were combined with an additional 3 g sample to give $\sim$8 g of powder. This was packed into an Al foil envelope and placed in an Al can. Two experiments were performed using the MARI direct geometry chopper instrument located at ISIS and also the Disk Chopper Spectrometer (DCS) at the NIST reactor source.  On MARI, a Gd chopper was used to obtain an incident energy E$_{i}$=40 meV and measurements were carried out between 5 and 350 K.   A further scan with E$_{i}$ = 150 meV was collected at 5 K to search for any higher energy excitations. To normalize these data on an absolute scale, the elastic incoherent scattering from a known Vanadium standard was used.  On DCS, an incident energy of 25 meV was used and the sample was enclosed in an Aluminum can with helium exchange gas.

\section{Rietveld Refinements}

Further details from Rietveld refinements (Ref. \onlinecite{Rietveld69:1}) using neutron powder data collected at 216 K and 2 K is provided. Refinements were carried out using a nuclear structure in space group $I$4/$mmm$ with La on 4$e$ site (${1\over2}$ ${1\over2}$ $z$), Fe on $4c$ site, Se on $4e$ site (0 0 $z$), O(1) on $4d$ site and O(2) on $2b$ site. The nuclear structural parameters are shown in Table \ref{table_refine}. 

\begin{table}[ht]
\caption{Refined Parameters}
\centering
\begin{tabular} {c c c}
\hline\hline
X & T=216 K & T= 2 K\\
\hline\hline
$a$ / \AA& 4.0792(2) & 4.0736(3) \\
$c$ / \AA & 18.563 & 18.515(2) \\
volume / \AA $^3$ & 308.89(4) & 307.25(6) \\
La $z$ & 0.1840(1) & 0.1838(1) \\
Se $z$ & 0.0964(2) & 0.0964(2) \\
R$_{wp}$ / \% & 5.165 & 5.383 \\
\hline
\label{table_refine}
\end{tabular}
\end{table}

The main text describes Rietveld refinements for the magnetic structure using neutron powder diffraction data collected at 2 K with magnetic Bragg peaks fitted by the 2-$k$ magnetic phase. This phase and the collinear phase give equivalent fits and those for the collinear phase are shown here in Figure \ref{collinear} $a-c$ for comparison.

\renewcommand{\thefigure}{S1}
\begin{figure}[t]
\includegraphics[width=0.95\linewidth,angle=0.0]{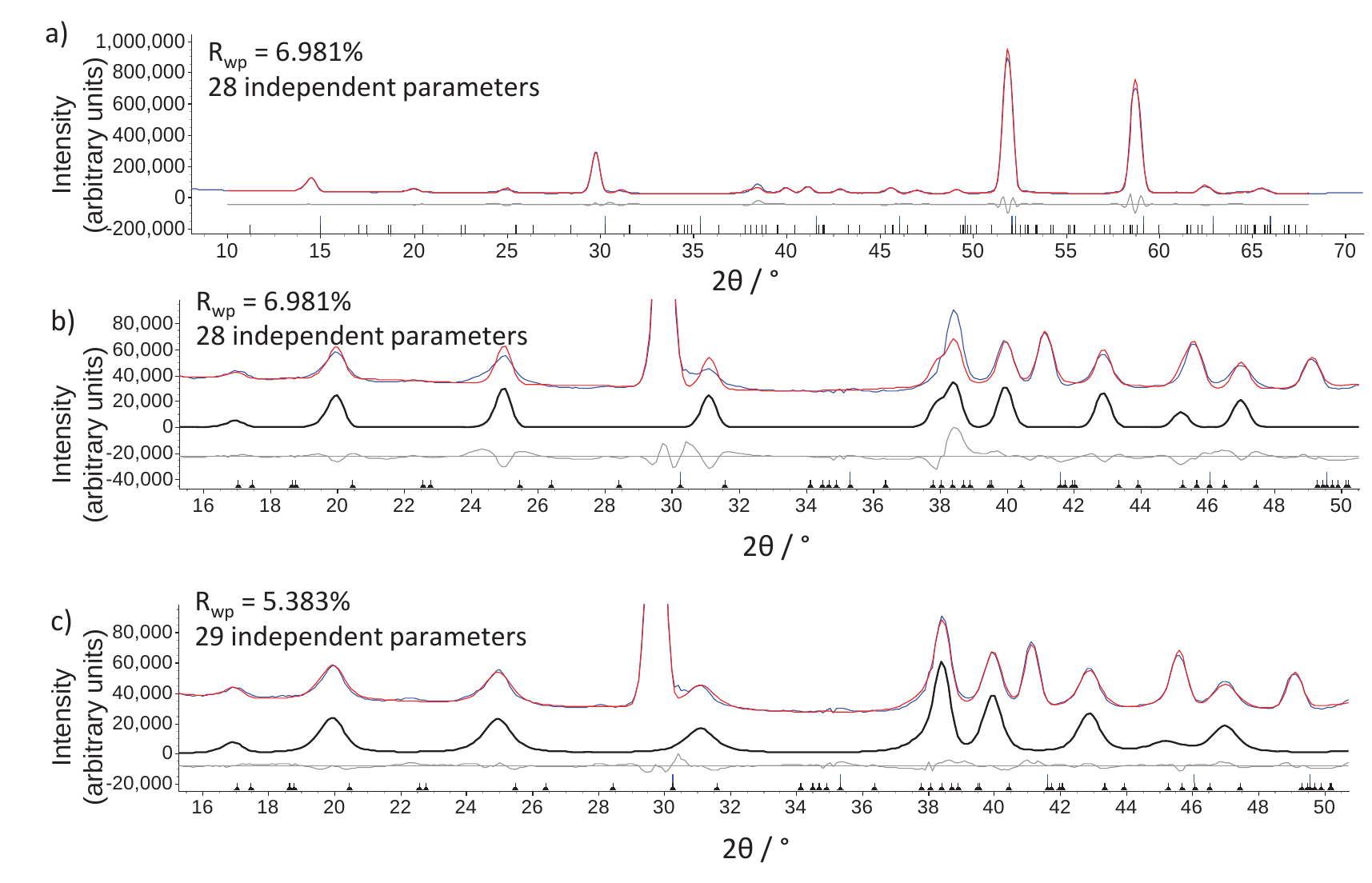}
\caption{[color online]  Rietveld refinement profiles (D20 with $\lambda$=2.41 \AA) for refinements with the collinear model showing a) wide 2$\theta$ range and b) showing narrow 2$\theta$ range to highlight magnetic Bragg reflections, with the same peak shape used for both magnetic and nuclear phases; c) refinement including antiphase boundaries applied to the magnetic phase. Observed, calculated and difference profiles are shown in blue and red (top) and grey (bottom), respectively.  The black tick marks indicate positions of possible magnetic Bragg peaks and the solid black line highlighs scattering from the magnetic-only phase.}
\label{collinear}
\end{figure}

\section{Warren Lineshape}

\renewcommand{\thefigure}{S2}
\begin{figure}[t]
\includegraphics[width=0.95\linewidth,angle=0.0]{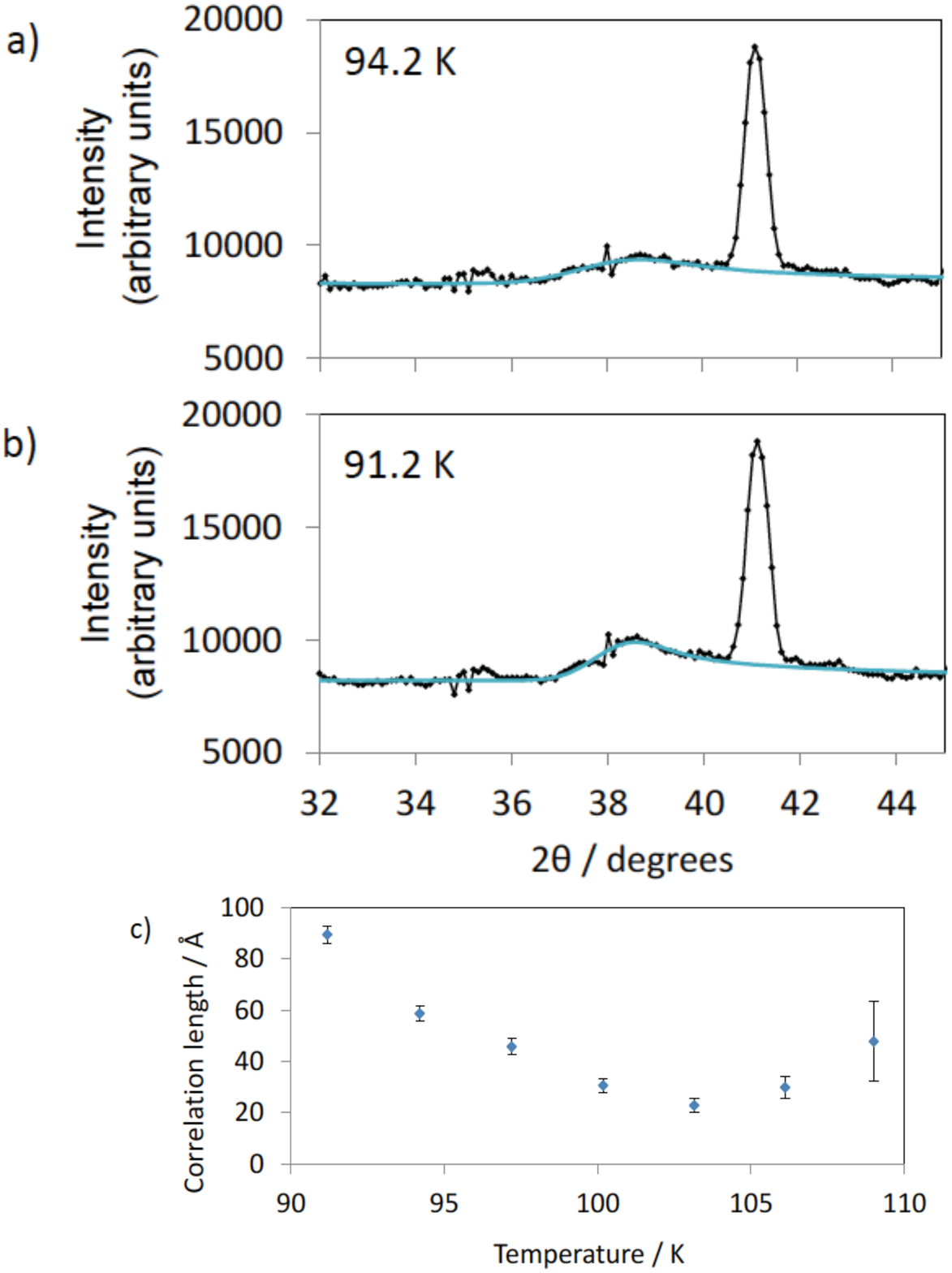}
\caption{[color online]  $a)-b)$ Examples of the Warren lineshape characterizing the two dimensional short-range correlations in La$_{2}$O$_{2}$Fe$_{2}$OSe$_{2}$.  The fit is given by the solid blue line.  $c)$ shows a plot of the correlation length extracted from simlar fits as a function of temperature.}
\label{Warren}
\end{figure}

Near the critical temperature, $T_{\mathrm{N}}\sim 89$ K, diffuse, two-dimensional, critical scattering was observed and fitted to the Warren lineshape (Ref. \onlinecite{Warren41:59}, Equation \ref{warren}),

\begin{eqnarray}
P_{2\theta}=K m F_{hk}^{2} {{1+\cos^{2}2\theta} \over {2 (\sin\theta)^{3/2}}} \left( {L} \over {\lambda \sqrt{\pi}}  \right) F(a)
\label{warren}
\end{eqnarray}

\noindent where $a=\left( {2L\sqrt{\pi}} \over {\lambda} \right) (\sin \theta - \sin \theta_{0})$ and $F(a)=\int_{0}^{\infty} \exp(-x^{2}-a^{2}) dx$, $K$ is the scale factor, $m$ is the mulitplicity of the reflection $(hk)$ centered at 2$\theta_{0}$, $F_{hk}$ is the structure factor (assumed to be constant over this narrow 2$\theta$ range) and $\lambda$ is the wavelength of radiation used.

Examples of Warren lineshapes are displayed in Fig. \ref{Warren} $a)$ and $b)$ and a plot of the fitted temperature dependent correlation length as a function of temperature is displayed in Fig. \ref{Warren} $c)$.  Two dimensional diffuse scattering was only observed over a narrow range near $T_{\mathrm{N}}$ reflecting the low levels of frustration in the magnetic structure as discussed in the main text. We note that this correlation length is different from the temperature independent magnetic correlation length corresponding to magnetic stacking faults along the $c$-axis which results in a broadening of the magnetic reflections at low temperatures.

\renewcommand{\thefigure}{S3}
\begin{figure*}[t]
\includegraphics[width=1.0\linewidth,angle=0.0]{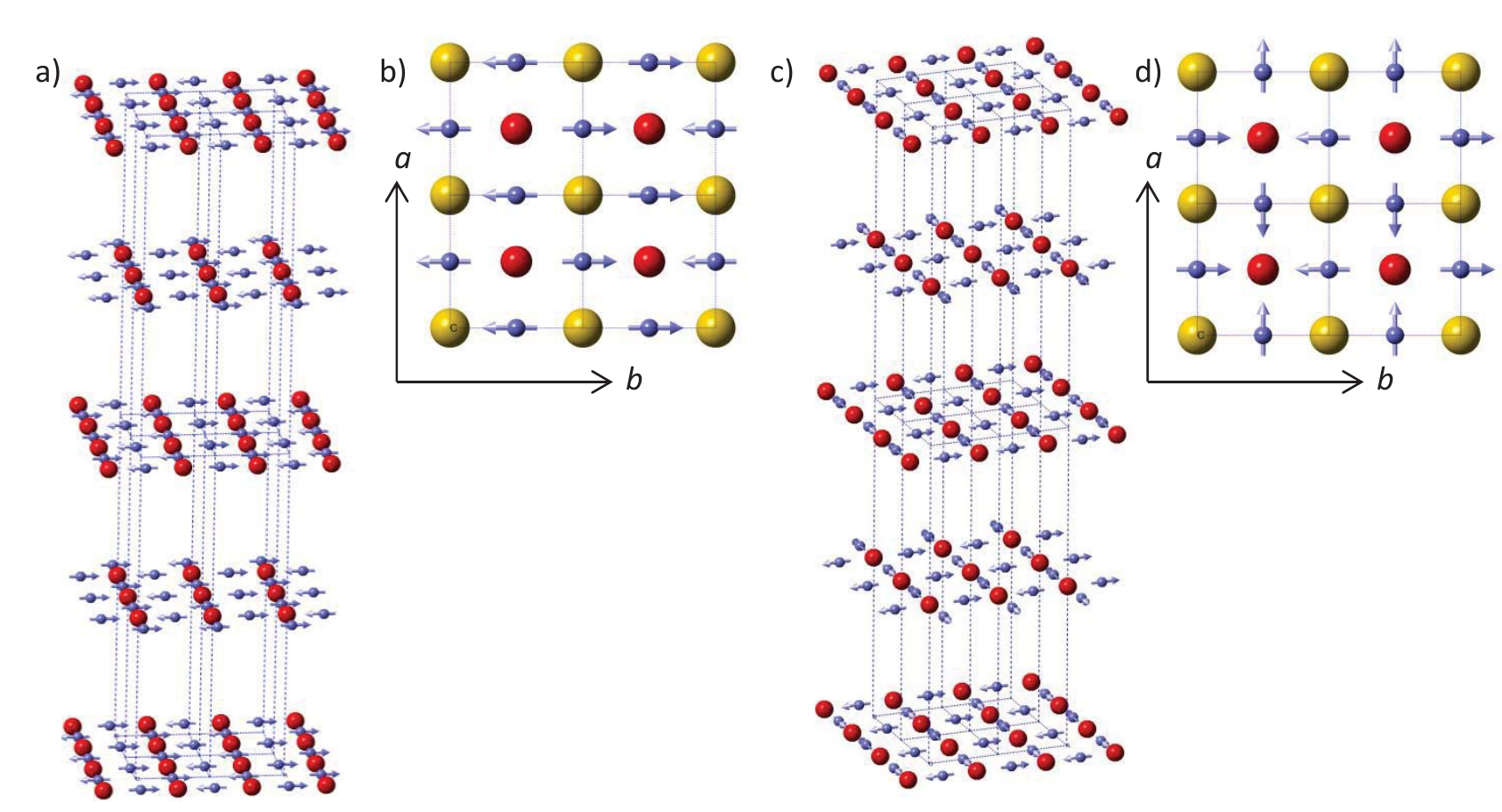}
\caption{[color online]   Candidate magnetic structures to describe the magnetic ordering in La$_{2}$O$_{2}$Fe$_{2}$OSe$_{2}$. $a)$ and $b)$ illustrate the collinear structure first proposed in Ref. \onlinecite{Free10:81}.  $c)$ and $d)$ show the 2-$k$ structure first proposed in Ref. \onlinecite{Fuwa10:22}.}
\label{large_structure}
\end{figure*}

\section{Candidate magnetic structures in La$_{2}$O$_{2}$Fe$_{2}$OSe$_{2}$ and relative ground state energies}

In the main text, it was pointed out that there have been two structures proposed to describe the magnetic diffraction patterns in La$_{2}$O$_{2}$Fe$_{2}$OSe$_{2}$ and related materials - the 2-$k$ model and the collinear model.  Detailed illustrations of two proposed magnetic structures are shown in Fig. \ref{large_structure} showing $a)$ collinear model with $b)$ view down long axis and $c)$ 2-$k$ model with $d)$ view down long axis, showing only O$^{2-}$ (red) and Fe$^{2+}$ (blue) ions for clarity, with Fe$^{2+}$ moments shown by blue arrows.  While the structure corresponds to a 2-$k$ model when based on the crystallographic unit cell, analysis using ISODISTORT~\cite{Campbell06:39} indicates that it can alternatively be described using a smaller $C$-centred monoclinic unit cell with magnetic space group $C_a$2/$m$ [number 12.64, with the basis (2, -2, 0) (2, 2, 0) (${1\over2}$ ${1\over2}$ 0) and origin (0, 0, 0)] with moments within the $ab$ plane, illustrated in Fig. \ref{monoclinic}).

\renewcommand{\thefigure}{S4}
\begin{figure}[b]
\includegraphics[width=0.5\linewidth,angle=0.0]{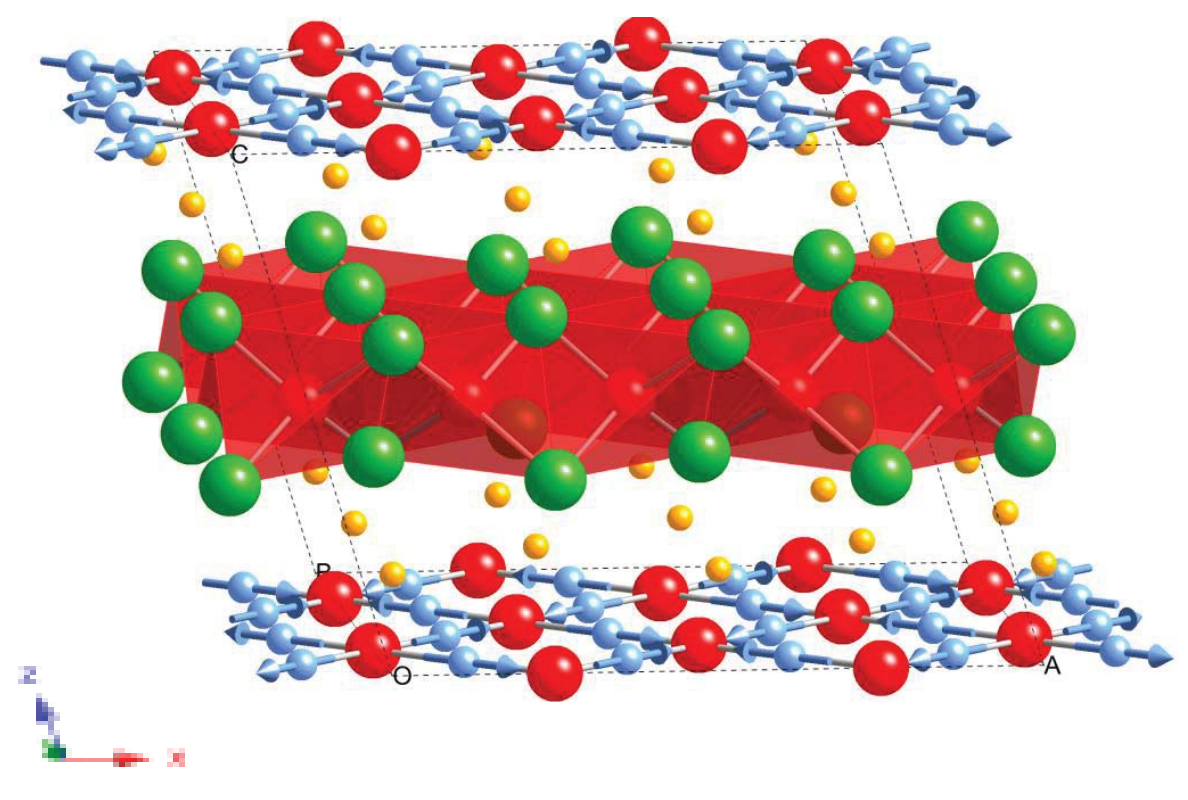}
\caption{[color online]  Illustration of the 2-$k$ magnetic model in the smaller $C_a$2/$m$ monoclinic space group with cell parameters $a_{m}$=$b_{m}$=2$\sqrt{2}$ $a_{n} \sim$11.5 \AA, $c_{m}=\sqrt{\left( {a_{n} \over \sqrt{2}} \right)^{2} +\left( {c_{n} \over 2} \right)^{2}} \sim 9.70$\AA, $\beta=90^{\circ}+\tan^{-1} \left( {a_{n}/\sqrt{2} \over c_{n}/2} \right)\sim107.3^{\circ}$ and cell volume $\sim$ 1230 \AA$^{3}$.}
\label{monoclinic}
\end{figure}

From our analysis of the magnetic structure and the spin excitations described in the main text, we believe the 2-$k$ structure is favored.  The structure was first proposed to describe the magnetic structure in  Nd$_{2}$O$_{2}$Fe$_{2}$OSe$_{2}$ and has more recently been applied to Sr$_{2}$F$_{2}$Fe$_{2}$OS$_{2}$.   As discussed in the main text, the structure can be viewed in terms of two interpenetrating lattices made up of single $k$ structures as illustrated in Fig \ref{2k_structure}. 

\renewcommand{\thefigure}{S5}
\begin{figure}[t]
\includegraphics[width=0.5\linewidth,angle=0.0]{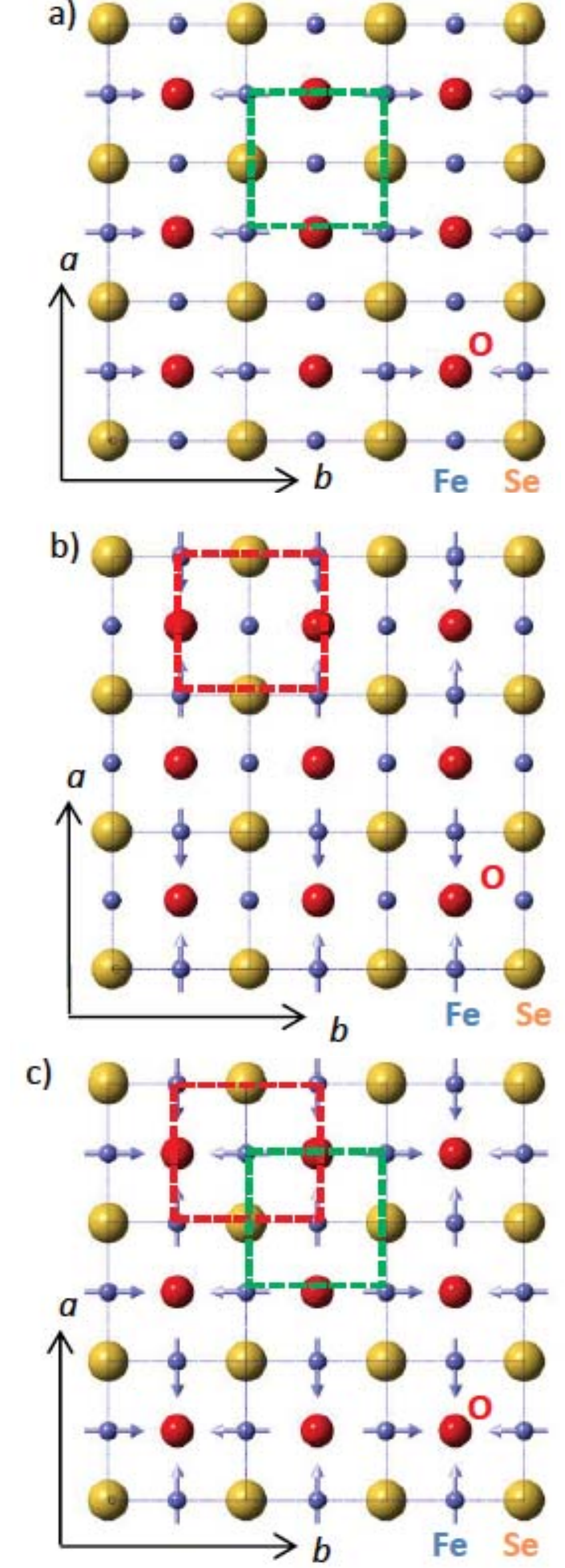}
\caption{[color online]   Schematic illustration of $k$ vectors a) $\vec{k}$=$(0, {1 \over 2}, {1\over 2})$  and b) $\vec{k}$=$({1 \over 2}, 0, {1\over 2})$ that together form the 2-$k$ structure, shown in c). The unit cell of the crystal structure is shown in blue and the interpenetrating sublattices are highlighted with green and red dashed lines. }
\label{2k_structure}
\end{figure}

While we have focussed the discussion in this paper around the collinear and 2-$k$ structures, several other magnetic structures have been proposed to fit related compounds.  Relative energies of the proposed magnetic structures based on experimental values for $J_{1}$, $J_{2}$ and $J_{2}'$ derived in the main text for the 2-$k$ structure are presented in Table \ref{relative_energies}.  The exchange constants were taken as  $J_{1}$ = -0.75 meV, $J_{2}$ = 0.10 meV, $J_{2}'$ = -1.00 meV and here we use the convention that a negative sign denotes AFM interaction to be consistent with Kabbour \textit{et al.}   For descriptions of structures FM, AF1, AF2 and AF3, see Kabbour \textit{et al.} (Ref. \onlinecite{Kabbour08:130}); for Wu stripe and Wu plane AF see Wu (Ref. \onlinecite{Wu10:82}); $\vec{k}$=$({1\over 2}, {1 \over 2}, 0)$ structure is that described for La$_{2}$O$_{2}$Co$_{2}$OSe$_{2}$, see Free and Evans (Ref. \onlinecite{Free10:81}) and Fuwa \textit{et al.} (Ref. \onlinecite{Fuwa10:22}).

\begin{center}
\begin{table*}[ht]
\caption{Spin exchange energies per chemical unit cell for the following spin states for La$_{2}$O$_{2}$Fe$_{2}$OSe$_{2}$:}
\begin{tabular} {c c c c}
\hline\hline
Structure & Ground State & Energy & comment\\
\hline\hline
FM & E$_{FM}$=4 $\times$(-8$J_{1}$-4$J_{2}$-4$J_{2}'$) & =38.4 meV & -- \\
AF1 & E$_{AF1}$=4 $\times$(8$J_{1}$-4$J_{2}$-4$J_{2}'$) & =-9.6 meV & Stabilized over FM by 48 meV \\
AF2 & E$_{AF2}$=4 $\times$(8$J_{1}$-4$J_{2}$-4$J_{2}'$) & =27.2 meV & Stabilized over FM by 11.2 meV \\
AF3 & E$_{AF3}$=4 $\times$(-4$J_{1}$-2$J_{2}$-4$J_{2}'$) & =19.2 meV & Stabilized over FM by 19.2 meV \\
Wu Stripe & E$_{Wu_s}$=4 $\times$(0$J_{1}$+4$J_{2}$+4$J_{2}'$) & =-14.4 meV & Stabilized over FM by 52.8 meV \\
Wu Plane AF & E$_{Wu_p}$=4 $\times$(0$J_{1}$-4$J_{2}$+4$J_{2}'$) & =-17.4 meV & Stabilized over FM by 56 meV \\
$\vec{k}$=$({1\over 2}, {1 \over 2}, 0)$ & E$_{ice}$=4 $\times$(0$J_{1}$+4$J_{2}$+4$J_{2}'$) & =-14.4 meV & Stabilized over FM by 52.8 meV \\
2-k & E$_{2-k}$=4 $\times$(0$J_{1}$-4$J_{2}$+4$J_{2}'$) & =-17.6 meV & Stabilized over FM by 56 meV \\
\hline
\label{relative_energies}
\end{tabular}
\end{table*}
\end{center}

\section{Zeroeth moment sum rule}

To compare our calculations with the experiment and to understand the excitation spectrum sensitive to the spin interactions, it is important to ensure that all spectral weight is accounted for and we have used the zeroeth moment sum rule to do this.  The zeroeth moment sum rule states that the integral of the measured intensity over all energy transfer (including the elastic, E=0, contribution) and momentum transfers is

\begin{eqnarray}
\tilde{I}(Q)={{\int_{-\infty}^{+\infty} dE \int_{0}^{Q} d^{3}q S(\vec{q},E)} \over {\int_{0}^{Q} d^{3}q}} = S(S+1).
\label{zeroeth_sum}
\end{eqnarray}

\noindent The integrated inelastic intensity contribution to the integral above gives I=3.2 (4), which is consistent with an ordered moment of 3.5 $\mu_{B}$ per Fe$^{2+}$ site.  Noting that the elastic line contributes a factor of $g^{2}\langle S_{z} \rangle^{2}$ (with $g=2$) gives $\langle S_{z} \rangle$=2.7.  The total integral above is therefore close to 6, as expected for $S$=2.

The analysis confirming this is presented in Fig. \ref{zero_moment} based upon a scan performed with E$_{i}$=150 meV to obtain a more complete detector coverage than allowed by lower incident energy data sets discussed in the main text.  The data is shown in  Fig. \ref{zero_moment}  $a)$ and the integral in Eqn. \ref{zeroeth_sum} is plotted in Fig. \ref{zero_moment} $b)$.  The initial rise at small momentum transfers is due to incomplete detector coverage and the large errorbars at larger momentum transfers are due to the decrease in intensity owing to the magnetic form factor.  The dashed line is the value of 3.5 quoted in the main text.

\renewcommand{\thefigure}{S6}
\begin{figure}[t]
\includegraphics[width=0.7\linewidth,angle=0.0]{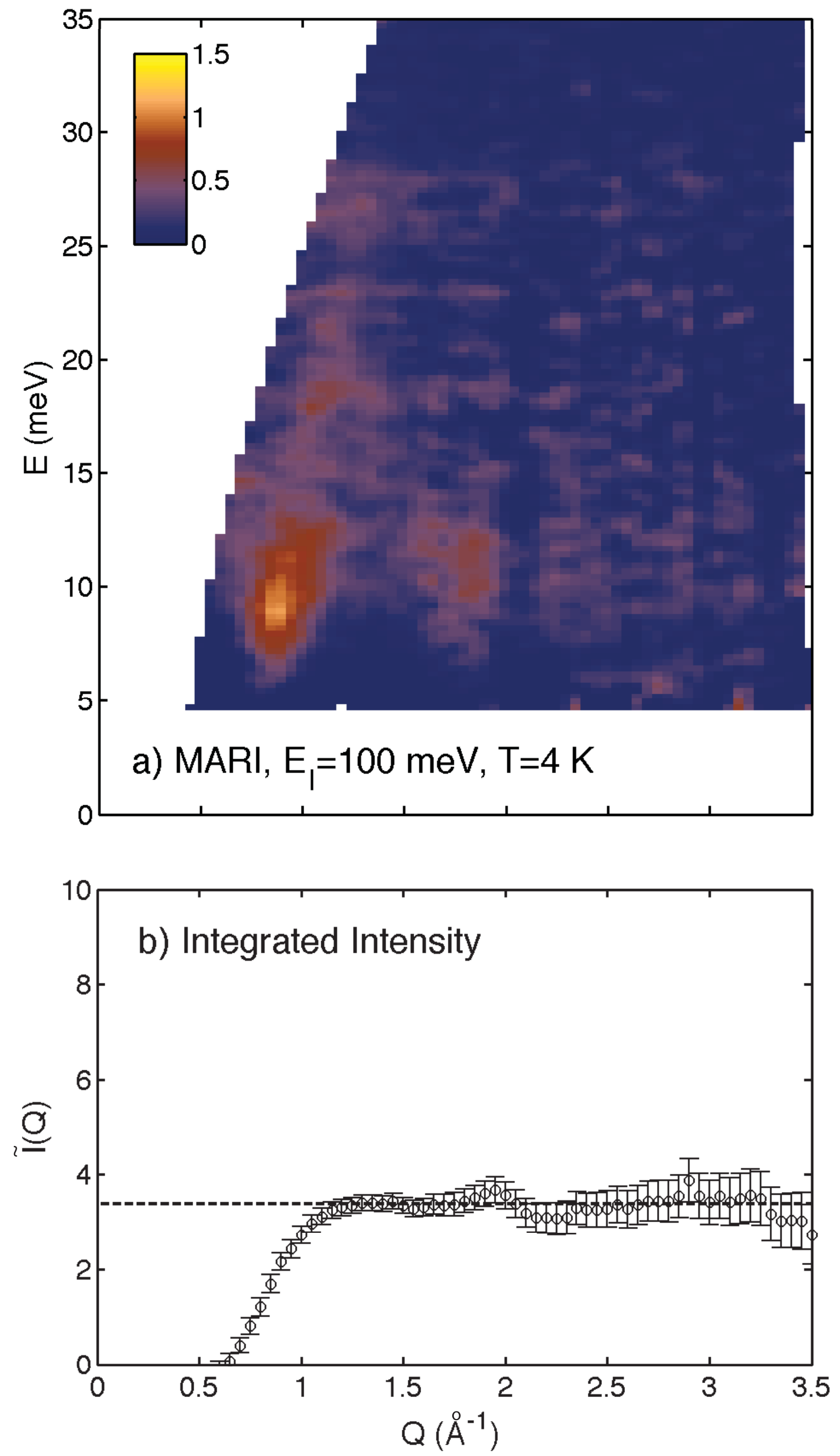}
\caption{[color online]   An analysis of the zeroeth moment sum rule. $a)$ illustrates a background corrected scan taken at with an E$_{i}$=100 meV and $b)$ shows the average integrated intensity as a function of a momentum transfer.  The dashed line is the average value used for the sum rule analysis described in the main text.}
\label{zero_moment}
\end{figure}

\section{First moment sum rule and single mode approximation}

As a first step towards understanding  the magnetic excitation spectrum, we utilized the single mode approximation combined with the first moment sum rule.  This method has been used previously to parametrize the excitations in low-dimensional and frustrated magnets.~\cite{Stock09:103,Hammar98:57}  In this framework, the measured structure factor can be written in terms of a momentum dependent part and a Dirac delta function in energy, 

\begin{eqnarray}
S(\vec{Q},E)=S(\vec{Q})\delta[E-\epsilon(\vec{Q})].
\label{zeroeth_sum}
\end{eqnarray}

\noindent The first-moment sum relates $S(\vec{Q})$ to the dispersion $\epsilon(\vec{Q})$,

\begin{eqnarray}
S(\vec{Q})=-{2\over 3} {1\over {\epsilon(\vec{Q})}}\sum_{\vec{d}} J_{d} \langle \vec{S}_{0} \cdot \vec{S}_{\vec{d}}\rangle [1-\cos(\vec{Q}\cdot\vec{d})].
\label{zeroeth_sum}
\end{eqnarray}

\noindent Here $\vec{d}$ is the bond vector connecting nearest neighbor spins with a superexchange $J_{d}$.  While this treatment is an approximation for our results on La$_{2}$O$_{2}$Fe$_{2}$OSe$_{2}$, a general result is that the powder averaged spectrum at the bottom of the dispersion at the gap edge is sensitive to the dimensionality of the exchange interaction.  Therefore, through a comparison of calcualtions with the lower edge of the measured powder averaged spectrum, the dimensionality of the spin exchange can be verified.

We have parameterized this in the main text of the paper in Fig.3 by calculating the neutron scattering intensity using the formalism above for wave vectors near the magnetic zone center.  We have described the dispersion of the single mode at small values of $\vec{q}$=(H-${1\over 2}$, K) as follows,

\begin{eqnarray}
\epsilon(\vec{Q})^{2}=\Delta^{2}+\alpha_{1}(H-{1\over2})^{2}+\alpha_{2}K^{2},
\label{sw_dispersion}
\end{eqnarray}

\noindent with the ratio ${\alpha_{1}\over\alpha_{2}}$ controlling the dimensionality.  The three dimensionally coupled limit (labelled 3D in Fig. 3 of the main text) was calculated by inserting a third $\alpha$ equal to $\alpha_{1,2}$.  The values of $\alpha$ were chosen so that the slope of the powder average excitations near the minimum of the dispersion curve agreed with experiment.  Regardless of this choice, the results illustrated in Fig. 3 are general.  The model calculations clearly show that the spin excitations are better described in terms of strong two dimensional interactions and is consistent with the 2D Ising universality class derived from a plot of the ordered moment as a function of temperature from the diffraction data.  

\section{Heuristic spin-wave analysis}

To parametrize the neutron inelastic data and extract exchange parameters that could be compared with electronic calculations and also compared to the magnetic structure, we have performed calculations considering a spin Hamiltonian consisting of Heisenberg exchange interactions and single-ion anisotropies.  The Hamiltonian used is described as,

\begin{eqnarray}
H_{stripe}=J_{1}\sum_{i,j}\vec{S}_{i}\cdot\vec{S}_{j}+J_{2}\sum_{i,j}\vec{S}_{i}\cdot\vec{S}_{j}+... \\
J_{2 '}\sum_{i,j}\vec{S}_{i}\cdot\vec{S}_{j} + D_{1}\sum_{i} S_{z,i}^{2} + D_{2}\sum_{i} (S_{y,i}^{2}-S_{x,i}^{2}). \nonumber
\label{hamiltonian_stripe}
\end{eqnarray}

\noindent The exchange constants $J_{1,2,2'}$ are schematically represented in Fig. 1 of the main text.  $J_{1}$ is the nearest neighbour exchange and $J_{2}$ and $J_{2}'$ are the next-nearest.  $J_{2}$ and $J_{2}'$ differ in that they are mediated by oxide and selenide anions, respectively. $D_{1}$ is an out of plane anisotropy which forces the moment direction to be within the $a-b$ plane.  The second term $D_{2}$ is input to stabilize a structure with moments selecting a preferred moment direction within the $a-b$ plane.   These anisotropy terms were used to stablize the magnetic structures so that exchange constants could be extracted.  Typically, the anisotropies were comparable to the extracted exchange constants which is unphysical.  This demonstrates that higher order terms in the Hamiltonian are likely needed to stablize the magnetic structures.

To calculate the spin excitations and the neutron cross section, we used the Holstein Primakoff operators.  The above expression for the spin Hamiltonian can be rewritten in terms of creation and annihilation operators in matrix form as follows,

\begin{eqnarray}
 {H \over {4S}}=\alpha^{\dagger} M \alpha\nonumber
\end{eqnarray}

$$
{H \over {4S}}=\left[a^{\dagger}_{q}  b^{\dagger}_{q} a_{-q}  b_{-q} \right] \left[\begin{array}{cccc}
A & C & D & F \\
C & B & F & E \\
D & F & A & C \\
F & E & C & B \\
\end{array}\right] \left[\begin{array}{c} a_{-q} \\ b_{-q} \\ a^{\dagger}_{q} \\ b^{\dagger}_{q}\end{array} \right]
$$

Calculations were performed for both the collinear and 2-$k$ magnetic ground states as discussed in the main text.

\noindent The matrix elements for the collinear model are as follows:

\begin{eqnarray}
A={J_{2}\over2}+J_{2}'\cos(\vec{q}\cdot\vec{b})+D_{1}-D_{2}
\label{A} \nonumber
\end{eqnarray}

\begin{eqnarray}
B={J_{2}'\over2}+J_{2}\cos(\vec{q}\cdot\vec{b})+D_{1}-D_{2}
\label{A} \nonumber
\end{eqnarray}

\begin{eqnarray}
C={J_{1}\over 2} \cos\left({{\vec{q}\cdot \vec{b}} \over 2}\right) \cos\left({{\vec{q}\cdot \vec{a}} \over 2}\right)
\label{A} \nonumber
\end{eqnarray}

\begin{eqnarray}
D=-J_{2}\cos(\vec{q}\cdot\vec{a})-{D_{1}\over 2}+{D_{2}\over 2}
\label{A} \nonumber
\end{eqnarray}

\begin{eqnarray}
E=-J_{2}'\cos(\vec{q}\cdot\vec{a})-{D_{1}\over 2}-{D_{2}\over 2}
\label{A} \nonumber
\end{eqnarray}

\begin{eqnarray}
F={J_{1}\over 2} \cos\left({{\vec{q}\cdot \vec{b}} \over 2}\right) \cos\left({{\vec{q}\cdot \vec{a}} \over 2}\right).
\label{A} \nonumber
\end{eqnarray}

\noindent The matrix elements for the 2-$k$ structure are as follows,

\begin{eqnarray}
A={J_{2}\cos(\vec{q}\cdot\vec{a})-2J_{2}-2J_{2}'+D_{1}-D_{2}}
\label{A} \nonumber
\end{eqnarray}

\begin{eqnarray}
B={J_{2}\cos(\vec{q}\cdot\vec{b})-2J_{2}-2J_{2}'+D_{1}-D_{2}}
\label{A} \nonumber
\end{eqnarray}

\begin{eqnarray}
C={J_{1} \cos\left({{\vec{q}\cdot \vec{b}} \over 2}\right) \cos\left({{\vec{q}\cdot \vec{a}} \over 2}\right)}
\label{A} \nonumber
\end{eqnarray}

\begin{eqnarray}
D=J_{2}'\cos(\vec{q}\cdot\vec{b})-{D_{1}\over 2}+{D_{2}\over 2}
\label{A} \nonumber
\end{eqnarray}

\begin{eqnarray}
E=J_{2}'\cos(\vec{q}\cdot\vec{a})-{D_{1}\over 2}-{D_{2}\over 2}
\label{A} \nonumber
\end{eqnarray}

\begin{eqnarray}
F={-J_{1} \cos\left({{\vec{q}\cdot \vec{b}} \over 2}\right) \cos\left({{\vec{q}\cdot \vec{a}} \over 2}\right)}.
\label{A} \nonumber
\end{eqnarray}

\noindent The matrix is Hermitian and follows several symmetry relations of the ground state illustrated in the two magnetic structures in question.  The energy positions were calculated from the eigenvalues of the matrix above and the neutron intensities were calculated from the eigenvectors.~\cite{Carlson04:70}

A combined analysis of the spin-waves and the diffraction indicates that the 2-$k$ structure is the most likely model.  The spin-wave analysis suggests antiferromagnetic $J_{1}$ and $J_{2 '}$ and ferromagnetic $J_{2}$ in contrast to the collinear structure which would require antiferromagnetic $J_{2}$ to reproduce the neutron inelastic results.  While a ferromagnetic $J_{2}$ is consistent with DFT calculations.  The superexchange through the 180$^{\circ}$ Fe-O-Fe path is expected to be strong and antiferromagnetic, while weak and ferromagnetic exchange is thought to occur via the $\sim$ 90$^{\circ}$ Fe-Se-Fe path.

\section{Curie-Weiss Temperature - A mean field description}

In this section we discuss the Curie-Weiss temperature and compare it to our heuristic model of the spin excitations.  This analysis assumes a mean-field description (Ref. \onlinecite{Smart:book}) and is likely only an approximation for La$_{2}$O$_{2}$Fe$_{2}$OSe$_{2}$ as a correct and full description of this point needs to include effects of magnetic frustration and low dimensionality.  The issue of dimensionality is particularly important given the presence of magnetic stacking faults evidenced by the Warren line shape and strong diffuse scattering observed in the magnetic diffraction pattern for this compound.  Nevertheless, while the discussion below in terms of mean-field theory is speculative, it is interesting in light of the spin-wave calculations and estimated values of the exchange constants.

Based on the exchange constants estimated from the calculations described above, a predicted value for the Curie-Weiss temperature can be derived based on the following formula where $S$ is the spin value, J$_{n}$ is the exchange interaction energy, and the sum is performed over all nearest neighbors,

\begin{eqnarray}
k_{B}\Theta_{CW}=-{1 \over 3} S(S+1) \sum_{n} J_{n}.
\end{eqnarray}

\noindent For the exchange interactions listed in Fig. 1 of the main text, this formula takes the following form for both the collinear and 2-$k$ magnetic structures,

\begin{eqnarray}
k_{B}\Theta_{CW}=-{1 \over 3} S(S+1) (4J_{1}+2J_{2}+2J_{2 '}).
\end{eqnarray}

\noindent Using this formula we estimate the expected $\Theta_{CW}$ temperature for the 2-$k$ magnetic structure to be $\sim$ -110 K and $\sim$ -75 K for the collinear variant.   Magnetic susceptibility data for La$_{2}$O$_{2}$Fe$_{2}$OSe$_{2}$ have been reported up to 300 K ~\cite{Mayer92:31} and further measurements to higher temperatures are required to investigate this material in the paramagnetic phase.  The fact that the Neel ordering temperature for the 2-$k$ structure in La$_{2}$O$_{2}$Fe$_{2}$OSe$_{2}$ is close to the calculated $\Theta_{CW}$ potentially reflects the lack of frustration in the system.  The small value predicted for the collinear structure is unphysical given the large degree of frustration in that structure discussed in the main text.  Therefore, a comparison of the predicted Curie-Weiss constant and the Neel temperature also points towards a 2-$k$ model being favored for La$_{2}$O$_{2}$Fe$_{2}$OSe$_{2}$.  Again, we emphasize that this discussion is based upon a mean-field description and does not account for frustrated exchange interactions or dimensionality.  The later is particularly important in La$_{2}$O$_{2}$Fe$_{2}$OSe$_{2}$ as evidenced by stacking faults along the $c$ axis.  A high degree of frustration might be expected to lower T$_{N}$. The observed T$_{N}$ (89.5(3) K) is similar to the calculated Weiss temperature ($\sim$100 K). Whilst this may reflect the low degree of frustration in the system, other factors such as the Ising anisotropy (which would tend to increase T$_{N}$) and the quasi-2D nature (which might lower T$_{N}$) must also be considered and the relative contributions of these factors is not known.


%